\documentstyle[prl,aps,floats,psfig,epsf,twocolumn]{revtex}
\tighten
\begin{document}
\draft
\title{Comment on "Superconducting Gap anisotropy vs. Doping Level in
High-T$_{c}$ Cuprates"}
\maketitle

In a recent paper Kendziora et al \cite{ken1} concluded that the superconducting
gap $\Delta(k)$ in overdoped 
Bi$_{2}$Sr$_{2}$CaCu$_{2}$O$_{8+\delta}$ (Bi-2212)
is isotropic. They analysed Raman spectra obtained 
in the $xy$(B$_{1g}$) scattering geometry, which samples regions
of the Fermi surface situated near the ($\pm$k,0),(0,$\pm$k) 
axes in reciprocal space, and in the $xx$(A$_{1g}$ + B$_{2g}$) geometry which
should yield \cite{che,che94,dev94} an approximate screened average over the 
Fermi surface. Their conclusion was based on the observation that pair
breaking peaks occurred at approximately the same frequency in both scattering
geometries, and that the normalized scattering intensity $(I_{s}/I_{n})$ at low
energies was strongly depleted in both geometries and appeared to be in 
reasonable agreement with theory. It must be noted however that their normalized
intensity included the Bose-Einstein factors at different temperatures and 
hence the resultant ratio exhibits an artificially enhanced depletion at low
energies and masks the low energy frequency dependance of the
response function. Furthermore, any comparison with theory is suspect since 
the normal state response function is unknown. Finally, since the relative
peak positions will also be very sensitive \cite{dev96} to the shape of the Fermi surface
and the form of the gap function,
they must be used with caution when comparing the symmetry of the superconducting
gap in crystals with significantly different doping levels. A consideration
of the low frequency behaviour of the $xy$ and $xx$ spectra
in conjunction with consideration of the B$_{2g}$ spectra implies the existence of an
anisotropic gap with nodes along the diagonals.

In Fig. 1 we present both the B$_{1g}$ and B$_{2g}$ response functions ($\chi$") 
which were obtained from an overdoped Bi-2212 crystal (T$_{c}$ = 55 K). 
Scattering in the B$_{2g}$ spectrum arises from regions of the Fermi surface
located near the diagonal directions in k-space and hence provides information
that is complementary to the B$_{1g}$ spectra. In both geometries it is clear
(Fig. 1) that (in the superconducting state) although scattering is suppressed 
for $\omega < 6k_{b}T_{c} (240 cm^{-1}$) it is still present and increases linearly with
$\omega$ in this frequency region. This linear behaviour is
incompatible with an isotropic s-wave gap. It is however consistent
with a model proposed by Devereaux \cite{dev96a} for a d-wave superconductor
with spin fluctuations and impurity scattering included.
The linear dependance of the B$_{2g}$ spectrum at low energies is
consistent \cite{che94,dev94,dev96} with the presence of nodes along the diagonal directions.
Hackl et al \cite{hac96a} also successfully interpreted data obtained from
an overdoped Bi-2212 crystal in terms of the disordered d-wave model.

\begin{figure}
\hskip2.cm
\caption[]{
The low energy B$_{1g}$ and B$_{2g}$ Raman spectra for an overdoped Bi-2212
crystal (T$_{c}$= 55 K) in the superconducting state (T = 15K). The straight
lines indicate the linear behaviour of the spectra for frequencies $\omega < \Delta_{max}$.
The intensity of the B$_{2g}$ spectra has been multiplied by a factor of 2.5
for comparison purposes.}
\label{fig1}
\end{figure} 

In conclusion the superconducting gap in overdoped Bi-2212 cannot be isotropic if
spectra from all scattering geometries are considered with thermal contributions
removed.
\vspace{0.1in}
\\{\it K. C. Hewitt, T. P. Devereaux, X. K. Chen, X-Z Wang, J. G. Naeini and J. C. Irwin \\
Department of Physics, Simon Fraser University,\\
Burnaby, B. C. V5A 1S6 Canada       
\vspace{0.1in}     
\\Airton Martin  \\
Instituto de Fisica "Gleb Wataghin", Universidade Estadual de Campinas,   \\
13083-970, Campinas, Sao Paulo, Brazil }    
\vspace{0.1in}
\\Received \today 

\bibliographystyle{prsty}

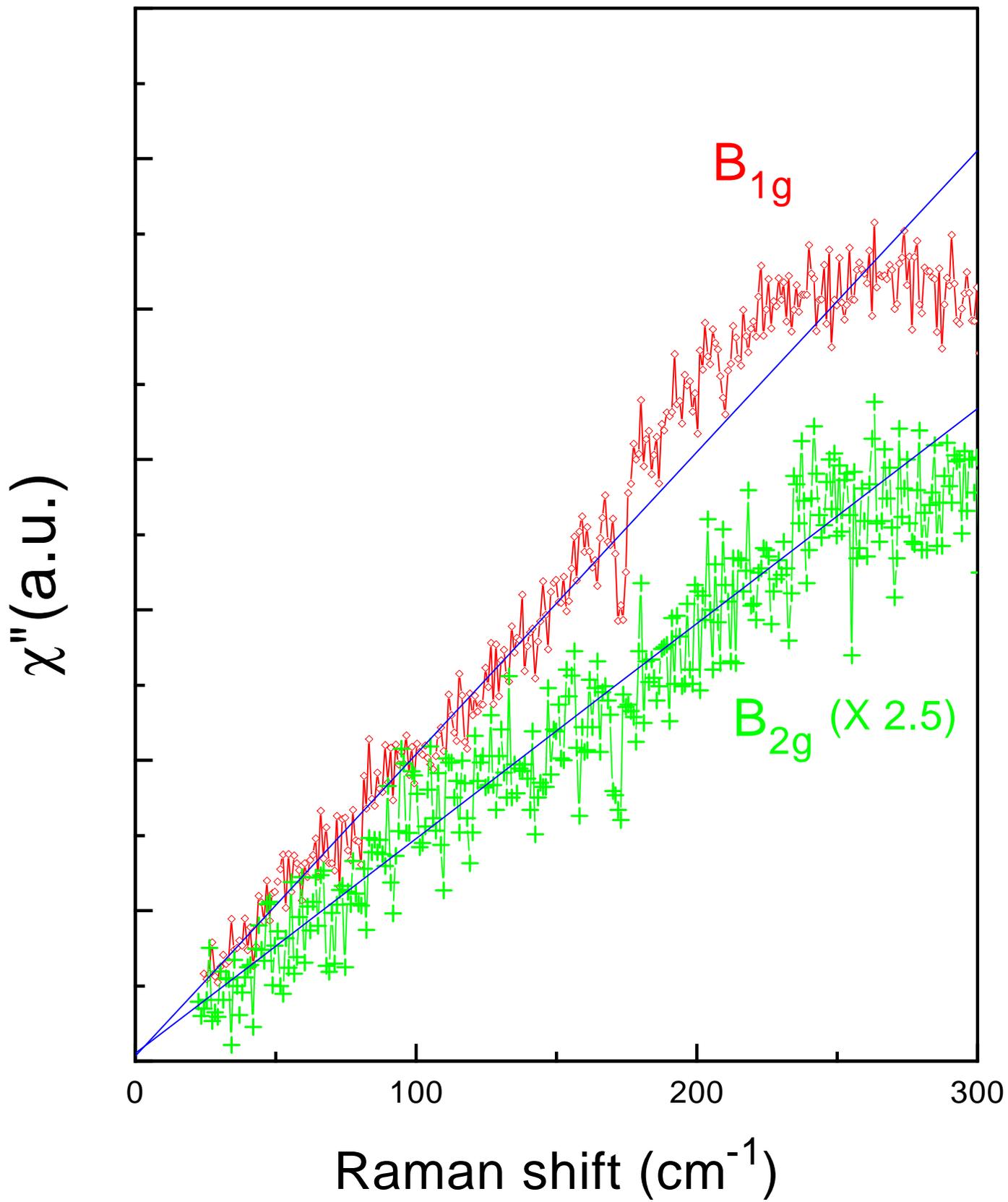
\end{document}